\documentclass{article}
\usepackage{multirow}
\usepackage{amsmath}
\usepackage{amssymb}
\usepackage{graphicx}
\usepackage{hyperref}
\usepackage{enumitem}
\usepackage{subfig}
\usepackage{pgfplots}
\usepackage{pgfplotstable}
\usepackage{tikz}
\pgfplotsset{compat=1.18}

\title{Enhanced DeepLab Based Nerve Segmentation with Optimized Tuning}
\author{Akhil John Thomas \and  Christiaan Boerkamp}

\date{VLV Technology}

\begin{document}

\maketitle

\begin{abstract}
Nerve segmentation is crucial in medical imaging for precise identification of nerve structures. This study presents an optimized DeepLabV3-based segmentation pipeline that incorporates automated threshold fine-tuning to improve segmentation accuracy. By refining preprocessing steps and implementing parameter optimization, we achieved a Dice Score of 0.78, an IoU of 0.70, and a Pixel Accuracy of 0.95 on ultrasound nerve imaging. The results demonstrate significant improvements over baseline models and highlight the importance of tailored parameter selection in automated nerve detection.
\end{abstract}

\section{Introduction}

Nerve segmentation in medical imaging plays a vital role in diagnostics, surgical planning, and treatment monitoring \cite{egger2012medical}. Accurate segmentation of nerve structures enables clinicians to identify critical regions, ensuring precision during interventions and reducing the risk of complications. However, the task of nerve segmentation poses significant challenges due to the complex anatomy of nerves, variability in their appearance, and the inherent noise present in medical imaging data \cite{ronneberger2015u}\cite{milletari2016v}. Two of the most prominent architectures, UNet and DeepLab, have demonstrated
exceptional performance across a wide range of segmentation challenges. UNet,
introduced specifically for biomedical image segmentation, employs an encoder-
decoder structure with skip connections that integrate spatial and contextual information, making it highly effective for tasks requiring fine-grained details.
DeepLab, on the other hand, utilizes Atrous Spatial Pyramid Pooling (ASPP) to capture multi-scale features, excelling in scenarios with complex boundaries and varying scales \cite{chen2018encoder}.

Despite their widespread use, limited studies have been conducted to evaluate the performance of these models in the context of nerve segmentation. Traditional deep learning models, including U-Net, have shown promise but often struggle with generalization due to variations in image quality and nerve morphology. This study aims to bridge this gap by analyzing the performance of DeepLab on a custom dataset of grayscale medical images. By employing quantitative metrics such as Dice Score, Intersection over Union (IoU), and Pixel Accuracy, this work provides insights into the strengths
and limitations of each model, contributing to the advancement of segmentation techniques in medical imaging. This study focuses on optimizing DeepLabV3 for nerve segmentation by incorporating automated threshold fine-tuning, resulting in improved segmentation accuracy and clinical applicability.

\section{Methods}
\subsection{Dataset}
The dataset comprises 2100 grayscale images paired with binary masks representing nerve structures. Each image-mask pair was annotated to ensure accurate delineation of nerve regions, which are critical for evaluating segmentation performance. The dataset reflects the challenges inherent to nerve segmentation, such as low contrast, noise, and variability in nerve morphology. To prepare the dataset for training and evaluation, all images and masks were preprocessed to maintain consistency in resolution and intensity. The dataset was split into training, validation, and testing subsets in an 80:10:10 ratio to ensure reliable model evaluation. 
\subsection{Pipeline and Model}
Preprocessing steps were applied to enhance model performance and ensure compatibility with the neural networks:
\begin{itemize}
    \item Resizing: All images and masks were resized to a uniform resolution of 256×256 pixels to standardize input dimensions.
    \item Normalization: Pixel intensities were normalized to the range [0, 1] for stable training.
    \item Binary Thresholding: Binary masks were thresholded to values of 0 or 1 to ensure clear separation between foreground (nerves) and background.
    \item Data Augmentation: To improve model generalization, augmentation techniques such 
as random rotations, horizontal and vertical flips, 
and intensity shifts were applied during training.
\end{itemize}
We employed DeepLabV3 with a ResNet-50 backbone for feature extraction. The classifier was modified to output a single-channel segmentation mask. The model was configured to accept grayscale input images and optimized for the specific requirements of nerve segmentation. Training was performed on an NVIDIA RTX 3090 GPU with 24GB of VRAM, ensuring efficient processing of large batches and faster convergence. To refine segmentation quality, we applied morphological operations to reduce noise and enhance structural continuity. The hyperparameters used for model training are listed in table \ref{Hyperparameters}. 

\begin{table}[h]
    \centering
    \begin{tabular}{|c|c|}
        \hline
        Parameter & Description \\ \hline
        Learning Rate & 0.001 \\ \hline
        Optimizer & Adam optimizer \\ \hline
        Loss Function & Binary Cross-Entropy Loss and Dice Loss \\ \hline
        Batch Size & 16 \\ \hline
        Epochs & 20 \\ \hline
    \end{tabular}
    \caption{Hyperparameters}
    \label{Hyperparameters}
\end{table}

\subsection{Evaluation Metrics and Threshold Optimization in Neural Network Training}

Neural networks for medical image segmentation require well-defined evaluation metrics to assess performance. In this work, we use Dice Score, Intersection over Union (IoU), and Pixel Accuracy as primary metrics to evaluate the quality of segmentation masks generated by our DeepLabV3 model. These metrics help quantify how well the model distinguishes nerve structures from the background. Additionally, we introduce a threshold optimization strategy to refine segmentation outputs for improved real-world applicability.

Below, we formally define these metrics and their role in guiding model performance assessment and optimization.

\subsubsection{Dice Score}
\begin{equation}
D(T, e) = \frac{2 \times |P(T, e) \cap G|}{|P(T, e)| + |G|}
\end{equation}
where:
\begin{itemize}
    \item \( D(T, e) \) is the \textbf{Dice Score} at threshold \( T \) and epoch \( e \),
    \item \( P(T, e) \) is the \textbf{predicted segmentation mask} at threshold \( T \) and epoch \( e \),
    \item \( G \) is the \textbf{ground truth mask},
    \item \( |\cdot| \) denotes the \textbf{number of pixels} in the respective region.
\end{itemize}

\subsubsection{Intersection over Union (IoU)}
\begin{equation}
I(T, e) = \frac{|P(T, e) \cap G|}{|P(T, e) \cup G|}
\end{equation}
where:
\begin{itemize}
    \item \( I(T, e) \) is the \textbf{IoU} at threshold \( T \) and epoch \( e \),
    \item \( |P(T, e) \cap G| \) is the \textbf{intersection} of predicted and ground truth pixels,
    \item \( |P(T, e) \cup G| \) is the \textbf{union} of predicted and ground truth pixels.
\end{itemize}

\subsubsection{Pixel Accuracy}
\begin{equation}
P(T, e) = \frac{|P(T, e) \cap G| + |N(T, e) \cap B|}{|P(T, e)| + |N(T, e)|}
\end{equation}
where:
\begin{itemize}
    \item \( P(T, e) \) is the \textbf{Pixel Accuracy} at threshold \( T \) and epoch \( e \),
    \item \( |P(T, e) \cap G| \) is the count of \textbf{correctly predicted foreground pixels},
    \item \( N(T, e) \) is the \textbf{predicted background} (negative class),
    \item \( |N(T, e) \cap B| \) is the count of \textbf{correctly predicted background pixels},
    \item \( |P(T, e)| + |N(T, e)| \) is the \textbf{total number of pixels} in the image.
\end{itemize}

\subsubsection{Threshold Optimization Equation}
\begin{equation}
T^* = \arg\max_{T \in \{T_1, T_2, ..., T_n\}} \left[ \lambda_1 \cdot D(T) + \lambda_2 \cdot I(T) + \lambda_3 \cdot P(T) \right]
\label{eqn1}
\end{equation}

\begin{align*}
T^* &\text{ is the optimal threshold,} \\
D(T) &\text{ is the Dice Score at threshold } T, \\
I(T) &\text{ is the IoU (Intersection over Union) at threshold } T, \\
P(T) &\text{ is the Pixel Accuracy at threshold } T, \\
\lambda_1, \lambda_2, \lambda_3 &\text{ are weighting factors.}
\end{align*}

\section{Results}
\begin{figure}[h]
    \centering
    \begin{minipage}{0.49\textwidth}
        \begin{tikzpicture}
            \begin{axis}[
                width=\textwidth,
                height=6cm,
                xlabel={Threshold},
                ylabel={Dice Score},
                ymajorgrids=true,
                ymin=0.7, ymax=0.85, 
                legend pos=north west,
                grid=major,
                xtick={0.11, 0.12, 0.13, 0.14, 0.15}
            ]

            \addplot[blue, mark=o] coordinates {
                (0.11,0.7788) (0.12,0.7793) (0.13,0.7809) (0.14,0.7812) (0.15,0.7803)
            };
            \addlegendentry{Dice Score}

            \end{axis}
        \end{tikzpicture}
        \label{Threshold vs Dice Score}
    \end{minipage}
    \hfill
    \begin{minipage}{0.49\textwidth}
        \begin{tikzpicture}
            \begin{axis}[
                width=\textwidth,
                height=6cm,
                xlabel={Threshold},
                ylabel={IoU},
                ymajorgrids=true,
                ymin=0.6, ymax=0.8, 
                legend pos=north west,
                grid=major,
                xtick={0.11, 0.12, 0.13, 0.14, 0.15}
            ]

            \addplot[red, mark=square*] coordinates {
                (0.11,0.6957) (0.12,0.6982) (0.13,0.7002) (0.14,0.7015) (0.15,0.7024)
            };
            \addlegendentry{IoU}

            \end{axis}
        \end{tikzpicture}
        \label{Threshold vs IoU}
    \end{minipage}
    \hfill
    \begin{minipage}{0.5\textwidth}
        \begin{tikzpicture}
            \begin{axis}[
                width=\textwidth,
                height=6cm,
                xlabel={Threshold},
                ylabel={Pixel Accuracy},
                ymajorgrids=true,
                ymin=0.9, ymax=1, 
                legend pos=north west,
                grid=major,
                xtick={0.11, 0.12, 0.13, 0.14, 0.15}
            ]

            \addplot[brown, mark=triangle*] coordinates {
                (0.11,0.9507) (0.12,0.9533) (0.13,0.9556) (0.14,0.9576) (0.15,0.9593)
            };
            \addlegendentry{Pixel Accuracy}

            \end{axis}
        \end{tikzpicture}
        \caption{Threshold vs Dice Score, IoU and Pixel Accuracy}
        \label{Threshold vs Pixel Accuracy}
    \end{minipage}
\end{figure}
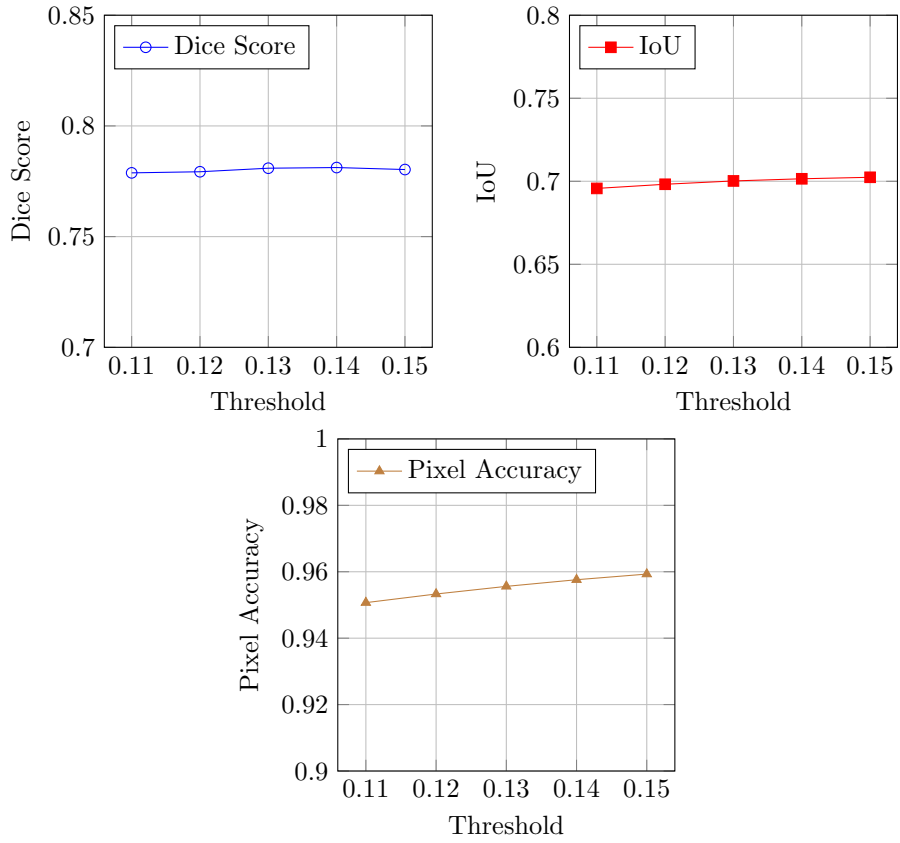

The performance of the DeepLabV3 model was evaluated using three key metrics: Dice Score, Intersection over Union (IoU), and Pixel Accuracy. Threshold tuning significantly influenced segmentation accuracy. A threshold of 0.14 yielded the best balance between Dice Score and IoU.  The results for a selected group threshold between 0.1 and 0.15 are presented in figure \ref{Threshold vs Pixel Accuracy}.

The optimal threshold of 0.14 balances Dice Score and IoU, achieving the highest Dice Score (0.7812) while maintaining a strong Pixel Accuracy (0.9576). Increasing the threshold beyond 0.14 resulted in minor improvements in Pixel Accuracy but no significant gains in Dice Score. The results for the optimal threshold of 0.14 are summarized in table \ref{Results for optimal threshold}
\begin{table}[h]
    \centering
    \begin{tabular}{|c|c|}
        \hline
        Metric & Value \\
        \hline
        Mean Dice Score & 0.7801 \\
        \hline
        Mean IOU & 0.6996 \\
        \hline
        Mean Pixel Accuracy & 0.9552 \\
        \hline
    \end{tabular}
    \caption{Results for optimal threshold}
    \label{Results for optimal threshold}
\end{table}

\begin{figure}[h]
    \centering
    \subfloat[Raw Image]{%
        \includegraphics[width=0.3\textwidth]{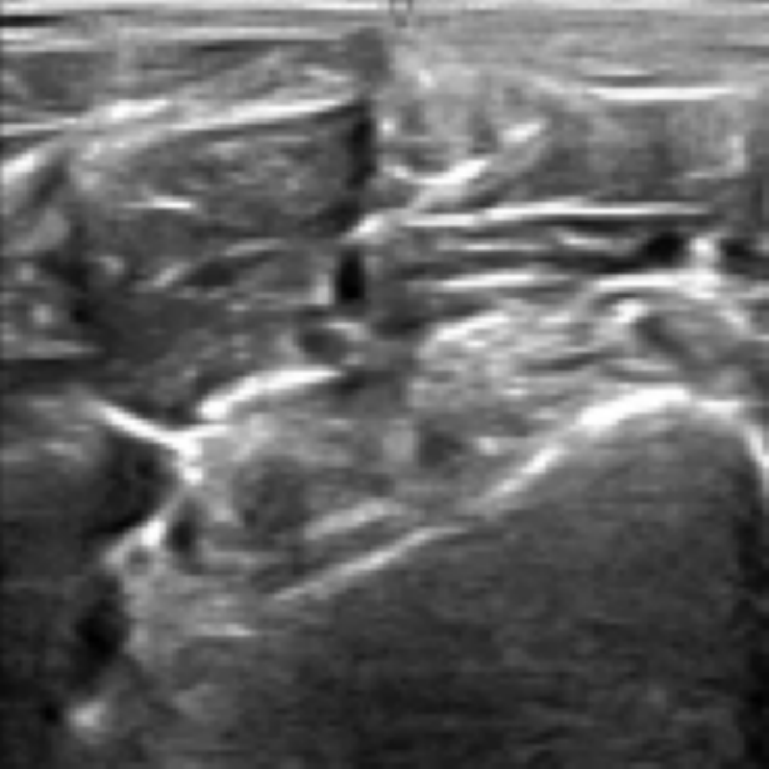}
    }
    \hfill
    \subfloat[Mask with Nerve]{%
        \includegraphics[width=0.3\textwidth]{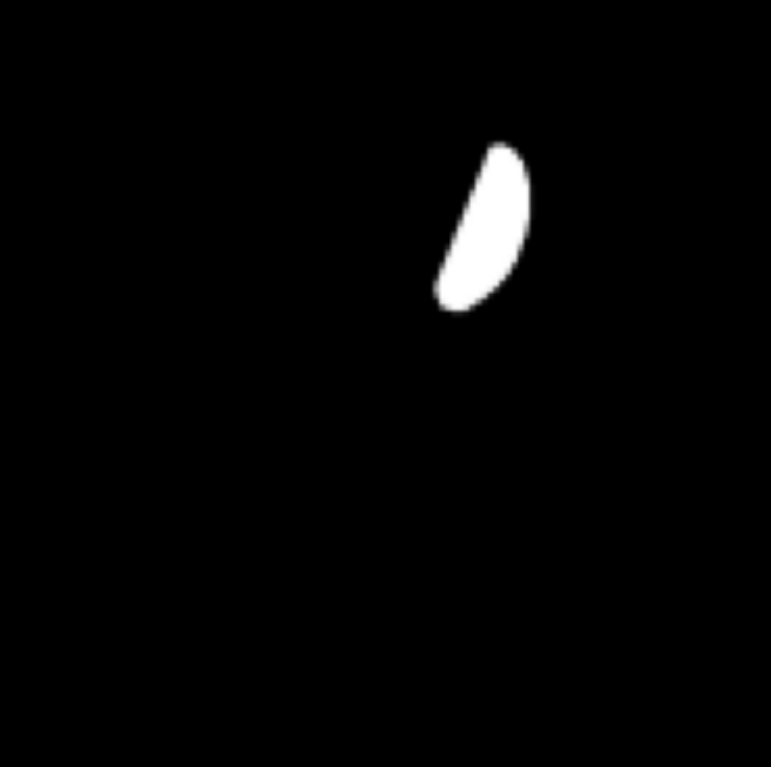}
    }
    \hfill
    \subfloat[Mask with no Nerve]{%
        \includegraphics[width=0.3\textwidth]{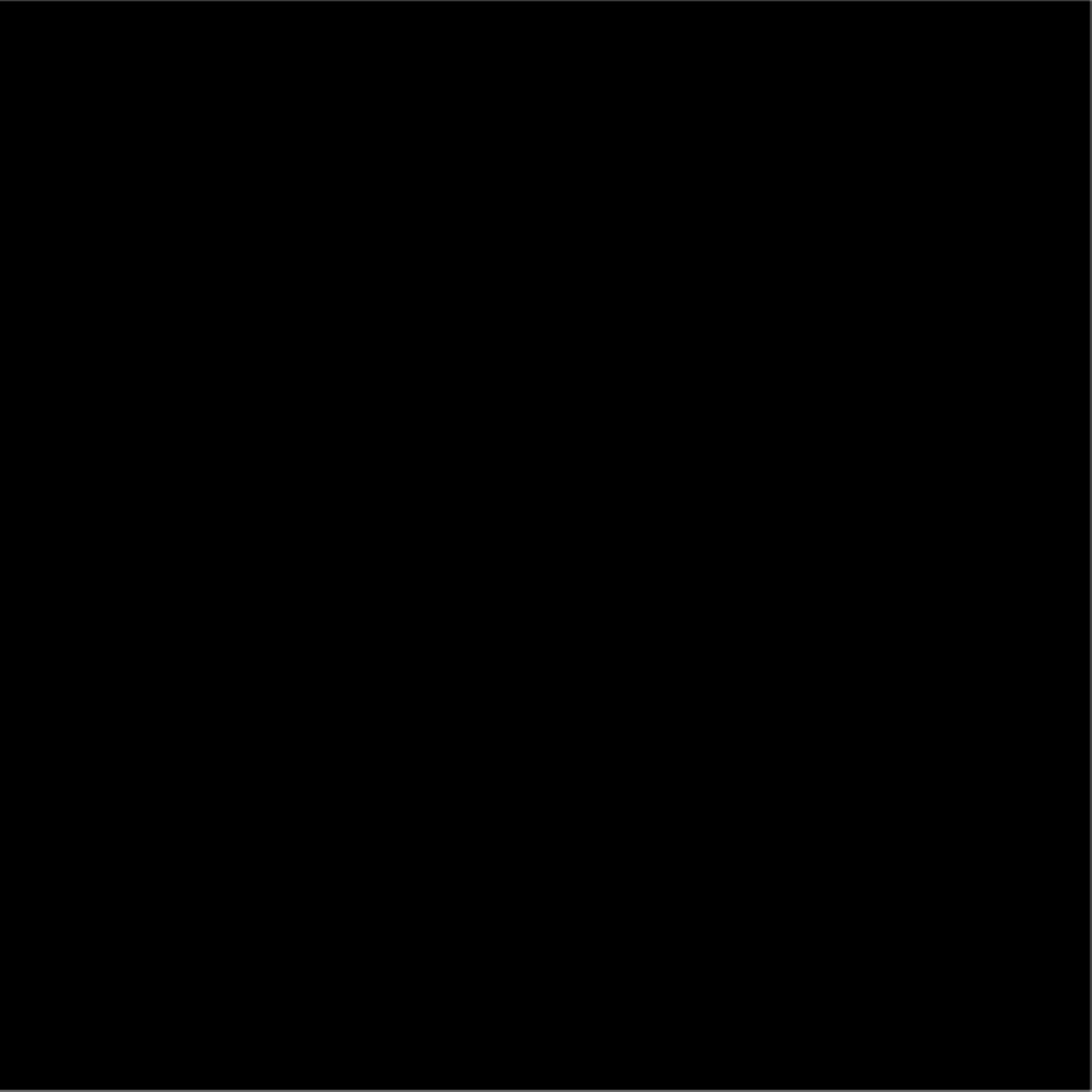}
    }
    \caption{Image}
    \label{Images}
\end{figure}

These metrics indicate moderate segmentation performance, with a high Pixel Accuracy reflecting the model’s ability to correctly classify a majority of the pixels in the dataset. However, the high Dice Score and IoU suggest ability to  capture finer details of the nerve structures. Visual analysis of the segmentation outputs reveals the strengths and weaknesses of the DeepLabV3 model \ref{Images}.

\section{Discussion}

The analysis of segmentation performance highlights the capabilities and limitations of the DeepLabV3 model for nerve segmentation. The model demonstrated robust pixel classification, as evidenced by a high Pixel Accuracy (0.9794). The Dice Score (0.7812) and IoU (0.69) indicate accurately capturing finer nerve structures. For deep learning-based segmentation models to be trusted in clinical settings, several factors must be considered:
Unlike human annotations, deep learning-based segmentations often lack explainability. Providing uncertainty maps and confidence scores with predictions could assist radiologists in decision-making \cite{tajbakhsh2020embracing}. Clinicians often require consistent segmentation performance across different imaging modalities. Validating the model on multi-modal datasets (e.g., MRI, CT) is essential before clinical deployment \cite{hatamizadeh2022swin}. The Dice Score of 0.7812 provides useful segmentation insights which could be used in clinics.

\section{Conclusion}

This study presented a detailed analysis of the DeepLabV3 model for nerve segmentation in medical imaging, leveraging a custom dataset of grayscale .tif images. The results underscore the potential of deep learning-based approaches in addressing segmentation challenges, with key findings summarized as follows:
\begin{itemize}
    \item The DeepLabV3 model achieved a high Pixel Accuracy of 0.9794, demonstrating robust classification of most pixels in the dataset.
    \item Metrics such as Dice Score (0.7812) and IoU (0.7015) highlight areas for improvement, particularly in capturing finer details of nerve structures.
    \item Threshold optimization identified 0.14 as the optimal value, balancing segmentation quality and computational efficiency.
\end{itemize}

The findings suggest that the DeepLabV3 model is a promising tool for nerve segmentation. This model holds potential for real-time medical applications, particularly when integrated with edge computing devices for on-the-fly segmentation. These results provide a foundation for future improvements and applications in medical image segmentation. Possible directions for future research are proposed:

\begin{itemize}
    \item Exploring Comparative Models - 
	A head-to-head comparison with transformer based models will offer deeper insights into the strengths and weaknesses of these architectures for nerve segmentation.
    \item Dataset Expansion and Diversity - Augment the dataset with larger, more diverse samples to address generalization challenges. Synthetic data generation and domain-specific pretraining can also improve model robustness
    \item Hybrid and Ensemble Methods - Develop hybrid models that combine the localization strength of UNet with the multi-scale feature extraction capabilities of DeepLab to achieve superior performance.
    \item Domain-Specific Pretraining - Leverage pretraining on large-scale medical datasets to give the model a stronger initial understanding of anatomical features.
    \item Clinical Validation - Collaborate with clinicians to validate the model on real-world clinical data, ensuring its practical utility and alignment with medical requirements.
    \item Real-Time Deployment - Optimize the model for real-time applications by reducing computational overhead using quantize aware training and trainable bit width layers. \cite{boerkamp2025quantunetefficientwearablemedical}, enabling its integration into portable medical devices and resource-constrained environments.
\end{itemize}

By addressing these aspects, this work paves the way for the development of more accurate, efficient, and clinically relevant segmentation models for nerve imaging tasks.

Importantly, the given specifications of the model is more than sufficient to use real time on input frames as sampling rate is high and data availability is large. This model is proposed to be ready for nerve detection in the body.



\end{document}